\documentclass[twocolumn,aps,prl,floatfix,superscriptaddress,showpacs]{revtex4}
\usepackage{graphicx}
\usepackage{bm}
\usepackage{epsfig}

\newcommand{\dsas}{\Delta_\mathrm{SAS}}
\newcommand{\dsasr}{\tilde\Delta_\mathrm{SAS}}
\newcommand{\dsw}{\Delta_\mathrm{sw}}

\begin{document}

\title{Dissipation and Tunnelling in Quantum Hall Bilayers}

\author{Robert L. Jack}
\affiliation{Theoretical Physics, University of Oxford, 
1 Keble Road, Oxford OX1 3NP, United Kingdom}
\author{Derek K. K. Lee}
\affiliation{Blackett Laboratory, Imperial College London, 
London SW7 2BW, United Kingdom}
\author{Nigel R. Cooper}
\affiliation{Cavendish Laboratory, Madingley Road, Cambridge CB3 0HE,
United Kingdom}

\begin{abstract}
We discuss the interplay between transport and intrinsic 
dissipation in quantum
Hall bilayers, within the framework of a simple thought experiment.
We compute, for the first time,
quantum corrections to the semiclassical dynamics of this system.
This allows us to re-interpret tunnelling measurements on these
systems.  We find a strong peak in the zero-temperature tunnelling
current that arises from the decay of Josephson-like oscillations into
incoherent charge fluctuations. In the presence of an in-plane field,
resonances in the tunnelling current develop an asymmetric lineshape.
\end{abstract}
\maketitle


Transport in quantum Hall bilayers has been the subject of much recent
interest\cite{qh_pers,Moon1995,Balents2001,bl_theory,Fertig2003,
Spielman2000on,Kellogg2004}. 
The bilayers consist of two closely-spaced parallel two-dimensional
electron layers in a double quantum well. If the Landau level fillings
of the layers are $\nu_1=\nu_2=1/2$, then Coulomb interactions
between the layers drive a transition to a ground state in which the 
bilayer as a whole exhibits the quantum Hall effect\cite{qh_pers}. 
This ground state is believed to have  
a broken U(1) symmetry\cite{bl_json}; there is a macroscopically coherent 
phase
associated with the electrons' layer degree of freedom. The ground
state can be viewed as an easy-plane ferromagnet\cite{Moon1995} or,
equivalently, as an excitonic condensate\cite{Fertig1989,Balents2001}.
There are also analogies with Josephson junctions\cite{bl_json}.

A series of remarkable
experiments\cite{Eisenstein92-94,Spielman2000on,Kellogg2004}
have been used to probe the internal degrees
of freedom of this strongly correlated system.
They show evidence for interlayer coherence and
the linearly dispersing Goldstone mode resulting from the broken
U(1) symmetry\cite{Spielman2000on}, and also, most recently, 
``excitonic'' superfluidity\cite{Kellogg2004}.

There still remain questions concerning the interlayer tunnelling 
spectrum of the bilayers.
One prominent feature in the $IV$ characteristic is a sharp peak in
the tunnelling current for small biases (between 10 and 100
$\mu$V)\cite{Spielman2000on}. At low temperatures, reducing the bias
leads to a sharp \emph{rise} in the tunnelling current. This increase
is cut off below 10 $\mu$V so that the current falls to zero at zero
bias. Existing theories for interlayer tunnelling are restricted to
the classical limit of the underlying spin model, and do not produce
this feature at the low temperatures relevant to experiment 
($k_BT\sim 2\mu \mathrm{eV}$). 

In this paper we study interlayer tunnelling within the framework
of a simple ``thought experiment''. We follow
the relaxation of an initial charge imbalance across the bilayer. 
Including quantum ($1/S$) corrections to the 
dynamics of the pseudospin model of the bilayer\cite{Moon1995},
we see that electron tunnelling across the bilayer generates
density waves. This quantum dissipative process
leads to a zero-temperature tunnelling current of the form $I\propto 1/V$
for a bias voltage $V$ above a threshold $V_0$; this is consistent with
experimental measurements. The long-wavelength
density fluctuations in the bilayer have an energy gap $\dsw$; in the absence
of disorder the
threshold $V_0$ corresponds to an energy of this order.
Introducing an in-plane magnetic field to our calculations
causes the small bias feature to develop into an asymmetric 
resonant peak in the tunnelling current, similar to
that reported in \cite{Spielman2000on}. Below the threshold $V_0$, we
find that the intrinsic dissipation causes any macroscopically
coherent charge oscillations to decay in time,
consistent with the absence of Josephson-like oscillations in
experiments. 

We work in the pseudospin picture of the bilayer\cite{Moon1995}. 
The charge imbalance on the bilayer is given by the $z$-component of the
magnetisation of the pseudospins. The system is a ferromagnet due to
Coulomb exchange. Since a $z$ component in the magnetisation
incurs a capacitative energy cost, 
the ferromagnet has easy-plane anisotropy.

We use a model of discrete spins on a lattice, beginning with a
spin-1/2 system in which each spin represents a local single-particle
state within the lowest Landau level. We work with a large-$S$ version
of this system. This large-$S$ generalisation may be treated
as a coarse-graining procedure.
The limit of large $S$ is the classical limit for the spin system.
We work with the Hamiltonian:
\begin{equation}
\frac{H}{2S} = - \frac{\rho_E}{2} \sum_{\langle ij \rangle} \vec{m}_i
\cdot \vec{m}_j + \frac{D}{4} \sum_i ( m_i^z )^2 - 
\frac{\Delta_\mathrm{SAS}}{2} \sum_i m^x_i 
\label{equ:H_spins}
\end{equation}
where $\vec{S}_i=S\vec{m}_i=S(m_i^x,m_i^y,m_i^z)$ is the spin operator
on site $i$ of a square lattice with spacing $c_0=\sqrt{2\pi} l_B$
where $l_B=(\hbar c/eB)^{1/2}$ is the magnetic length.  The 
exchange $\rho_E$ and the strength of the on-site repulsion
$D$ were derived from microscopic considerations by Moon \emph{et
al.}\cite{Moon1995}. 
The tunnelling between the layers enters the problem through
$\dsas$: the splitting of the ``bonding'' and ``anti-bonding''
single-particle states in the double well.
In our thought experiment, a gate is used to control
the charge imbalance on the bilayer: this adds a term $H_V = - SV
\sum_i m^z_i$ to the Hamiltonian.
Typical values for the model parameters in physical bilayer systems
are $l_B\simeq20\mathrm{nm}$, 
$\dsas\simeq90\mu\mathrm{K}$, $\rho_E\simeq0.5\mathrm{K}$, $D\simeq
30\mathrm{K}$.

We consider the properties of the ferromagnetic phase of this model,
which is believed to describe the experimental systems. In particular,
the observation of a linearly dispersing peak in the tunnelling
conductance\cite{Spielman2000on} indicates that the experimentally
accessible 
state has ferromagnetic order --- the peak arises from the
Goldstone mode of the system. We therefore ignore the 
quantum disordered phase that exists for $D/\rho_E S^2 \gg 1$ 
\cite{Chubukov1994}. 

\begin{figure}
\epsfig{file=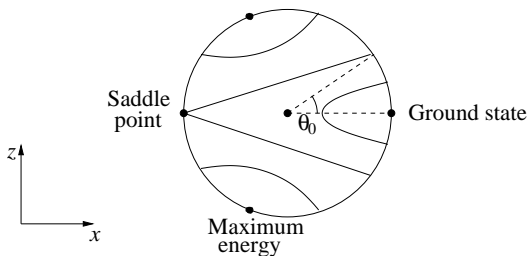,width=0.8\columnwidth}
\caption{Mean field (classical) spin trajectories are equipotentials
on the surface of the spin sphere. }
\label{fig:q0_spins}
\end{figure}

We will now outline the dynamics of the classical ferromagnetic system
before we discuss the quantum effects which are responsible for the
existence of a dissipative tunnelling current.  In the classical
system, a spatially uniform spin configuration will remain uniform
forever.  The total magnetisation
precesses along a trajectory of constant energy. We sketch the trajectories in
Fig.~\ref{fig:q0_spins} for $\dsas \ll D$ which is the physically
relevant regime.
 
Near the ($x$-polarised) ground state,
the precession frequency is given by $\dsw/\hbar$ where $\dsw=
[\dsas ( \dsas + D )]^{1/2}$ is the energy gap for spinwave
excitations (density waves of charge imbalance
across the bilayer).  As the magnetisation precesses along these
trajectories, the charge imbalance on the bilayer oscillates around
zero.  Far away from the ground state, the spin precesses around 
one of the two maximal energy states which lie close to the 
$S^z$ axis.
This yields a Josephson-like alternating current $I\simeq
e\dsas \cos (eVt/\hbar)$ where $V$ is the voltage across the bilayer
due to capacitative charging.  We stress that this is valid only for
large charge imbalance (large $V$). 

Trajectories through the saddle point at $\vec{m}=(-1,0,0)$
mark the boundary between oscillations around the ground state and 
oscillations around the maximal energy states. 
The saddle point trajectory crosses the $xz$-plane
at three points,
as shown in Fig.~\ref{fig:q0_spins}. The angle $\theta_0$ satisfies
$\cos\theta_0=1-2(\dsas/D)$. It corresponds to a voltage 
difference of  $V_0 = 2\dsw/e$ across the
layers.


In our thought experiment, we imagine using a gate to
induce a uniform charge imbalance on the bilayer. In the spin picture,
the magnetisation is tilted out of the easy ($xy$) plane. 
The bias is
then instantaneously removed and the bilayer finds itself in a highly
excited state.

The resulting behaviour depends crucially
on whether the initial gate voltage $V$ is above or 
below the saddle point value $V_0$. 
We treat the two regimes in separate perturbation theories.
In both calculations, we expand around
the classical limit using spinwave
theory in a $1/S$ expansion. The leading terms in
this expansion result in the leading terms in the $I$-$V$ relation.

We begin by considering the situation for initial energies above the saddle 
point ($V>V_0$). When the gate voltage is removed, the charge 
imbalance results in a potential difference
across the bilayer equal to the original gate voltage $V$. 
As discussed above, this causes an alternating current with a 
frequency $(eV/\hbar)$. In this regime, we can treat the tunnelling
$\dsas$ perturbatively.


The quantum spin system is distinct from the classical one in that
spatially uniform oscillations do not persist --- the tunnelling term breaks
the global spin rotation symmetry so that long-wavelength modes are no longer
protected from decay by Goldstone's theorem.  
The uniform mode decays by transferring energy into spin waves
with finite wavevectors. The magnetisation falls to a lower trajectory on the
spin sphere, corresponding to a net transfer of charge across the bilayer.
We can therefore obtain the d.c. tunnelling current by
calculating the rate of this dissipative process.

It is straightforward to derive a bosonic spinwave theory for our
model.  We use the Holstein-Primakov representation:
$S^x_j+iS^y_j=(2S-a^\dag_j
a_j)^{1/2} a_j$, $S^z_j = S-a^\dag_j a_j$.  Expanding around
$m^z=(eV/D)=\sin\theta$, the
quadratic part of the Hamiltonian is then easily diagonalised in the
Fourier basis to give
\begin{equation}
H^{(0)}_{\theta>\theta_0} = (D\sin\theta) \delta n_{q=0}
  + \sum_{\bm{q}\neq0} \varepsilon_{\bm{q}} \alpha_{\bm{q}}^\dag \alpha_{\bm{q}}
\end{equation}
where the $\alpha_{\bm{q}}$ are bosonic operators
describing the spinwave modes. The spinwave dispersion is given by
$\varepsilon_{\bm{q}} = [\rho_E \gamma(\bm{q}) ( D + \rho_E
\gamma(\bm{q}))]^{1/2}$ 
where $\gamma(\bm{q})=4-2\cos(q_x c_0) - 2\cos(q_y c_0)$. 
The dispersion is
linear and gapless at small $q$. (Tunnelling should produce a small 
energy gap but this does not affect the perturbative calculation we describe
here.) 
The spinwave velocity is $v=l_B (2\pi D\rho_E)^{1/2}$.
Observe that the $q=0$ mode has been singled out in the Hamiltonian, and
its energy is not given by the long wavelength limit of $\varepsilon_{\bm{q}}$.
The quanta of this mode carry $S^z=1$ while spin waves 
with finite wavevector have $S^z=0$.

\begin{figure}
\epsfig{file=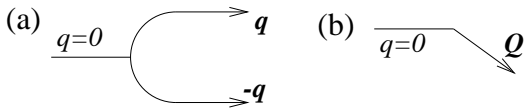, width=0.8\columnwidth}
\caption{Decay processes for $V>V_0$. (a) dominant process for Q=0: 
$\varepsilon_{\bm{q}}+\varepsilon_{-\bm{q}}=eV$.
(b) Decay process at finite $Q$: $\varepsilon_{\bm{Q}}=eV$. }
\label{fig:diag_asp}
\end{figure}

Starting from an initial state $|i\rangle$, the energy dissipation 
rate is given by 
\begin{equation}
\Gamma = \partial_t \Big< i \Big| e^{iHt} \sum_{\bm{q}\neq0} 
\varepsilon_{\bm{q}} \alpha^\dag_{\bm{q}} \alpha_{\bm{q}} e^{-iHt}
\Big| i \Big>
\end{equation}
The dissipation arises from the destruction of one quantum in the  
$q=0$ mode, and the
generation of multiple spinwaves during tunnelling across the
bilayer.  To leading order in $1/S$, a pair of spin waves 
with opposite momenta is excited
(Fig.~\ref{fig:diag_asp}a).  The relevant vertex is:
\begin{equation}
H^{(1)}_{\theta>\theta_0} = (-\dsasr/8) 
e^{-i\phi_{q=0}}
\sum_{\bm{q}} \gamma_{2,\bm{q}} \alpha^\dag_{\bm{q}} \alpha^\dag_{-\bm{q}}  + 
\hbox{h.c.}  
\end{equation}
in which $\dsasr = \dsas \exp({-\sqrt{D/\rho_E S^2}})$ is a
renormalised tunnelling amplitude, and 
$\gamma_{2,\bm{q}} = \cos\theta [ (u_{\bm{q}}+v_{\bm{q}})^2 +
2\sin\theta\sec^2\theta- (u_{\bm{q}}-v_{\bm{q}})^2\sec^4\theta ] $
where $u_{\bm{q}}$ and $v_{\bm{q}}$ are coherence factors:
$(u_{\bm{q}}+v_{\bm{q}})^2=(D+\rho_E\gamma(\bm{q}))/\varepsilon_{\bm{q}}$
with $u_{\bm{q}}^2- v_{\bm{q}}^2 =1$. The $\theta$-dependence arises
because the $x$-component of the pseudospin depends on its angle 
with the easy plane, $\theta$, as well as on the azimuthal
angle $\phi$. This dependence is weak for 
$\theta\ll 1$ when the charge imbalance is small
compared to the Landau level filling. 

\begin{figure}
\epsfig{file=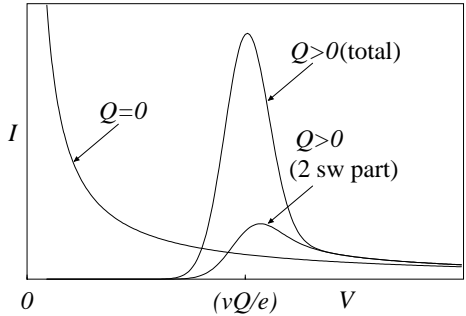,width=0.7\columnwidth}
\caption{Plots of the current $I$
(arbitrary units)
at $Q=0$ and at finite $Q=0.5 (4\pi\rho_E S/v)$ so that
$vQ\sim 150\mu\mathrm{V}$. The two spinwave contribution (2sw) to
the current gives rise to the asymmetric peak, as shown.
The contributions at finite $Q$ have been broadened by a Gaussian
distribution for $Q$ of width $\sigma_Q=0.1 (4\pi\rho_E S/v)$.}
\label{fig:iv}
\end{figure}

We now calculate the power dissipation
$\Gamma$ for a given initial voltage $V$; the steady-state tunnelling
current density at a bias $V$ is 
$I=\Gamma/VL^2 S$. The result is:
\begin{equation}
\frac{\Gamma_{\theta>\theta_0}}{L^2 S} = I_{\theta>\theta_0}V =
\frac{D \dsasr^2}{32 \pi \rho_E l_B^2\hbar S} 
\left[ 1 + X(\theta) \right]^2
\label{equ:I_asp}
\end{equation}
where $X(\theta)=(\sec\theta-1)(\sec^2\theta-2\sec\theta-1)$ is small
for a small charge imbalance ($\theta\ll 1$).
Thus, the power dissipation into spinwave pairs is
\emph{independent} of the voltage, and so the tunnelling current has a $1/V$
divergence. This should be cut off at low bias when the bias $V$ becomes
comparable to $V_0$ and perturbation theory breaks down. 
This contribution to the tunnelling current is 
significant at low temperatures: it is our main result,
consistent with the experimental observation\cite{Spielman2000on} 
of a peak at a bias
close to the spinwave gap. Its contribution to the tunnelling current
density is shown in Fig.~\ref{fig:iv}. 

We note that the region of the experimental sample 
in which the tunnelling current flows remains an open 
question\cite{Balents2001}.
Estimating the power dissipated in the experiments\cite{Spielman2000on}
to be $10^{-16}W$, and neglecting renormalisation of the tunnelling 
(so $\tilde\dsas=\dsas$), we obtain an estimate of $50\mu\mathrm{m}^2$
for this area. 
This estimate is increased if we use a renormalised $\dsas$.
The result is consistent with tunnelling taking place
near the contacts to the bilayer, rather than over its entire area. 
In the experimental data of \cite{Spielman2000on}, the current decays
more slowly than $1/V$ away from the resonance, so that the power increases
with increasing applied voltage.
We attribute this increase to other 
dissipative channels, due to disorder or finite temperature. 
Processes at higher order in $1/S$ also affect the current in this way
(see (\ref{equ:I_BR}) below), but are not large enough to explain the 
differences between theory and experiment.  

It should be noted that the dissipation rate (\ref{equ:I_asp}) is
averaged over the period of the Josephson oscillations. As with
other theories in which the tunnelling is treated 
perturbatively\cite{Balents2001,bl_theory},
we also expect an
oscillatory a.c.~component with frequency $eV/\hbar$. 

At this point, we make contact with previous calculations of the
tunnelling current in \cite{Balents2001}. If we ignore the weak
$\theta$-dependence of the vertex factor, the dynamics depend only on
the azimuthal angle $\phi$ of the pseudospin. This is the same
theory as in \cite{Balents2001}. However, that work gives a
\emph{vanishing} tunnelling current at zero temperature.
Our calculation is different from that one since we
include quantum corrections involving multi-spinwave processes.

More quantitatively, the current is
calculated in \cite{Balents2001} from $I_\phi \sim \int d^2{\bm{r}} dt
\exp(ieVt/\hbar -G_Q)$ where the quantum propagator $G_Q$ is, in our
notation:
\begin{equation}
G_Q(\bm{r},t) = -i (v\hbar/2\pi S \rho_E) 
\left[ (vt-ia)^2 - r^2 \right]^{-1/2}\,.
\end{equation}
We have explicitly included a lattice cutoff, $a\sim
l_B$, that reflects the finite bandwidth of the spin waves. We recall that
our parameters satisfy $(\hbar v/\rho_E S l_B) < 1$ to avoid
quantum disordering. 
We can therefore perform the integral order by order in $(\hbar v/\rho_E S
a)$, finding:
\begin{equation}
I_\phi =  \frac{ \exp(-eVa/\hbar v)}{V} 
\frac{D\dsasr^2} {16\pi \hbar l_B^2\rho_E} 
\frac{\mathrm{d}^2}{\mathrm{d}\Omega^2} I_0(2\sqrt{\Omega})
\label{equ:I_BR}
\end{equation}
where $\Omega=eV/4\pi S \rho_E$ 
and $I_0$ is a modified
Bessel function. The leading terms
in the voltage $V$ are also the leading ones in $1/S$, justifying the
expansion about the classical limit.

In an expansion in $eV/\rho_E S$, the leading term in
eq.~(\ref{equ:I_BR}) coincides with eq.~(\ref{equ:I_asp}) if we ignore
$X(\theta)$. In fact, the
$n$th term in the expansion arises from the decay of a single quantum
of the $q=0$ mode into $n+2$ finite-momentum spinwaves.
Thus, the zero-temperature tunnelling
current arises from multi-spinwave processes. These contributions have
been ignored in \cite{Balents2001} although they are, in principle,
contained in the framework of that paper.

We now generalise to the case in which a magnetic field is applied in
the plane of the bilayer. In that case, the tunnelling term in the
Hamiltonian acquires a spatial dependence: $\dsas m^x\to \dsas [m^x
\cos(Qx) + m^y\sin(Qx)]$, where $Q=(e B_y d/\hbar c)^{1/2}$ is the
wavelength associated with the in-plane field $B_y$. The spacing, $d$,
between the two electron layers is approximately $30\mathrm{nm}$.

The tunnelling vertex now involves a change of momentum $Q$ in the
$x$-direction.  A new process appears: a single $q=0$ quantum may decay
into a single spinwave (Fig.~\ref{fig:diag_asp}b). The
contribution from this process leads to a $\delta$ function peak in
the dissipation. The two-spinwave process contributes to the
dissipation rate only for $eV > vQ$. The leading contributions to the
current density are
\begin{eqnarray}
I_{\theta>\theta_0,Q} &=&
\frac{D \dsasr^2}{16\pi\hbar l_B^2 } \times
\Big[  2\pi V^{-1} \delta(eV-vQ) +
 \nonumber\\
& & \phantom{hi} \;\frac{1}{2 \rho_E S} \frac{e}{\sqrt{(eV)^2-(vQ)^2}}
\Theta(eV-vQ) \Big]
\end{eqnarray}
where $\Theta(x)$ is the step function. 
The peak at $eV=vQ$ is asymmetric. 
The lineshape is controlled by the
multi-spinwave processes. For $eV<vQ$, there is no available spinwave
channel for dissipation and so there is no tunnelling current: at
larger voltages, the current decays as a power law. 
This is shown in  Fig.~\ref{fig:iv}, 
in which we
have broadened the delta function by introducing low-momentum
scattering with 
momentum spread $\sigma_Q$. This may arise from disorder or thermal
effects. In the absence of these
effects, the delta function will be broadened
by the intrinsic lifetimes of the final states (for
example, by further tunnelling or spinwave/spinwave scattering).


The treatment thus far has been valid only for
voltages greater than $V_0=2\dsw/e$. We now discuss charging
voltages smaller than this value. 
In this regime, our thought experiment does not result in a 
steady tunnelling current.
Instead, the
current and charge imbalance oscillate coherently around zero with
frequency
$\dsw/\hbar$. We now evaluate the rate at which these oscillations
decay.

We treat the parameter $\theta$ perturbatively in this regime. 
In the Holstein-Primakov representation, the
quadratic part of the Hamiltonian takes the form:
\begin{equation}
H_{\theta<\theta_0}^{(0)} = \sum_{\bm{q}} \omega_{\bm{q}} \beta^\dag_{\bm{q}}
\beta_{\bm{q}}
\end{equation}
where $\omega_q = \left[ ( \dsas + \rho_E \gamma(\bm{q}) )( D + \dsas
+ \rho_E \gamma(\bm{q}) ) \right]^{1/2}$ and $\beta_{\bm{q}}$
annihilates a spinwave with wavevector $\bm{q}$.  

The initial state of our thought experiment has a charge imbalance. This
corresponds to a condensate of 
bosons with $q$=0. 
These bosons have the small concentration
$n_0/S=\theta^2 [1+(D/\dsas)]^{1/2}/4\pi
l_B^2$.
The simplest
dissipative process involves four $q$=0 quanta annihilating to form a
pair of spinwaves with finite momenta. 
We defer details to a later paper\cite{Jack}:
the dissipation rate is
\begin{equation}
\frac{\Gamma_{\theta<\theta_0}}{L^2} = 
\left(\frac{\theta}{\theta_0}\right)^8\!\!
  \frac{\dsas^3 \dsw^2 }{ 512\pi \rho_E D^2 l_B^2\hbar}
 f\left(\frac{\dsas}{D}\right)^2 
\end{equation}
where $f(x)=(1+x)^{-4}[3-4x-8x^2-x\sqrt{1+16x+16x^2}]$ is
approximately equal to 3 since ${\dsas}\ll D$. This
dissipation corresponds a decay of the density of $q=0$ modes: $\dot n
= -\Gamma(\theta)/4\hbar\dsw L^2$. Since the density $n$ is proportional to
$\theta^2$, we see that $d\theta^2/dt \sim -\theta^8$. 
The amplitude $A$ of the oscillations
in the tunnelling current is proportional to $\sin\theta$, and so
it decays in time as a power law: $A(t) \sim 1/t^{1/6}$. 

The dissipation is very weak at small initial angles partly due to the
kinematic constraint for a 4-spinwave collision. One might expect
stronger dissipation in the presence of disorder or inelastic
scattering, \emph{e.g.}, with phonons or charged quasiparticles.
Nevertheless, the multi-spinwave processes provide an
\emph{intrinsic} damping mechanism that has not previously been
identified.

We note that this treatment does not lead to a state with a direct
tunnelling current at $V<V_0$.  Therefore we cannot address the
lineshape of the zero-bias peak in $\mathrm{d}I/\mathrm{d}V$. It is
worth noting that this feature has sharpened with improved
experimental conditions. It may be controlled by low-energy
states introduced by disorder or by effects arising from coupling to
external leads. An approach similar to \cite{Fertig2003} may be
needed.

In conclusion, we find that the zero-temperature tunnelling current
has a strong $1/V$ peak arising from the decoherence of Josephson-like
oscillations by the generation of electron density fluctuations in the
bilayer. We also predict a non-trivial lineshape for the 
dispersing feature that appears in the presence of an in-plane field.
These effects resemble features reported in
experiments\cite{Spielman2000on}.

We acknowledge the financial support of EPSRC (in part
through GR/S61263/01) and of the Royal Society.


\end{document}